\documentclass[12pt]{article}
\usepackage{amssymb,amsmath}
\usepackage{graphicx}
\usepackage{epsfig}
\usepackage{epstopdf}
\usepackage{latexsym}
\usepackage{psfrag}
\usepackage{subfigure}
\usepackage{booktabs}
\usepackage{bbm}

\usepackage[toc]{appendix}

\usepackage{color}
\usepackage{datetime}
\usepackage[
      colorlinks=false,
      linkcolor=darkblue,  
      urlcolor=blue,    
      filecolor=blue,     
      citecolor=red,
linktocpage=true,
      pdfstartview=FitV,
      bookmarksopen=true    
      ]{hyperref}

\ifpdf
\fi

\numberwithin{equation}{section}

\renewcommand{\theequation}{\arabic{section}.\arabic{equation}}

\def\coeff#1#2{\relax{\textstyle {#1 \over #2}}\displaystyle}

\def\IR{\mathbb{R}}

\newcommand{\omg}{\omega}

\newcommand{\bq}{\begin{equation}}
\newcommand{\eq}{\end{equation}}
\newcommand{\ra}{\rightarrow}
\newcommand{\Ra}{\Rightarrow}

\definecolor{cardinal}{rgb}{0.6,0,0}
\definecolor{darkgreen}{rgb}{0,0.5,0}
\definecolor{golden}{rgb}{0.92, 0.7, 0}
\definecolor{midnight}{rgb}{0, 0, 0.5}
\definecolor{darkblue}{rgb}{0.2, 0, 0.8}

\begin{document}  

\begin{titlepage}
 
\bigskip
\bigskip
\bigskip
\bigskip
\begin{center} 
{\Large \bf    Bubbling the Newly Grown Black Ring Hair}

\bigskip
\bigskip 

{\bf Orestis Vasilakis \\ }
\bigskip

Department of Physics and Astronomy \\
University of Southern California \\
Los Angeles, CA 90089, USA  \\
\bigskip
vasilaki@usc.edu\\
\end{center}

\begin{abstract}
 
New families of BPS black ring solutions with four electric and four dipole magnetic charges have recently been explicitly constructed and uplifted to M-theory. These solutions were found to belong to a CFT with central charge different compared to the one of the STU model. Because of their importance to AdS/CFT, here we give the microstate description of these geometries in terms of topological bubbles and supertubes.  The fourth charge results in an additional flux through the topological cycles that resolve the brane singularities. The analog of these solutions in the IIB frame yield a generalized regular supertube with three electric charges and one dipole charge.   Direct comparison is also made with the previously-known bubbled geometries.


\end{abstract}

\end{titlepage}



\section{Introduction}

Finding an effective description of the microscopic degrees of freedom of black holes at strong gravitational coupling has been a major program in black hole physics the last years. An exact enumeration of states has been done for the two charge geometries, where smooth supertube solutions were matched to states in the D1-D5 CFT \cite{Mathur:2005zp,Skenderis:2008qn}. \\

For systems with three charges, a significant step forward was made in \cite{Bena:2004de} were the nonlinear supergravity equations were reduced to a sequentially linear system of equations. The 1/8-BPS solutions were described in M-theory frame which, after compactification on $T^6$, can be truncated to five dimensional ungauged  supergravity coupled to two vector multiplets. The fascinating idea presented in \cite{Bena:2005va,Berglund:2005vb} (for a review see \cite{Bena:2007kg}) was that one can resolve the singularity and allow for a semiclassical description of microstates within the supergravity framework, by allowing the four dimensional base metric to be ambipolar, i.e. change sign from $(+,+,+,+)$ signature to $(-,-,-,-)$, while the full five dimensional metric retains its original signature $(-,+,+,+,+)$. Then the singularity is resolved by a geometric transition, where the brane charges are dissolved within magnetic fluxes threading topological cycles. By using Gibbons-Hawking metrics as a base space large families of examples of multi-centered ambipolar geometries have been constructed  \cite{Bena:2007kg,Bena:2005va,Bena:2006is,Bena:2006kb}. The solutions are regular provided that certain integrability equations or ``bubble equations", are satisfied.\\

A general description of a black rings in $N=2$ supegravity with n vector multiplets has been given in \cite{Elvang:2004ds,Gauntlett:2004qy}.  In \cite{Giusto:2012gt} the authors, motivated by the CFT analysis of \cite{Giusto:2011fy}, present an M-theory description of a BPS black ring which upon reduction to $T^6$ gives  five dimensional ungauged  supergravity coupled to \textit{three} vector multiplets. Thus the solution has more hair and, although it has four electric and four dipole magnetic charges, it is still 1/8-BPS. The similarity of the equations and the black ring solution between \cite{Giusto:2012gt} and \cite{Bena:2007kg} makes it natural to provide a microstate description by using ambipolar Gibbons-Hawking metrics. As it was mentioned in \cite{Giusto:2012gt}, these microstates are dual to states of CFT with central charge different from the one of the D1-D5 CFT which is dual to the STU model. Finding the subset of microstates within the current formalism, that are dual to D1-D5 CFT might provide a hint for the additional required states to account for the black ring entropy. In section 2 we summarize the supergravity description of \cite{Giusto:2012gt}. In section 3  we construct microstate geometries with the same asymptotic structure at infinity as the black rings and the bubble equations and the asymptotic charges are derived. A regular supertube solution with three electric and one dipole magnetic charge  that seems to exist in this framework is examined. The solution although singular in five dimensions can be smooth in IIB frame if certain conditions are satisfied. Final remarks are given in section 4 and a few details about Gibbons-Hawking metrics may be found in the appendix.

\section{The hairier BPS solution}

\subsection{General Form of the solution}

Let us review the supergravity solution of \cite{Giusto:2012gt} with the addition of the gauge transformations that leave the solution invariant.\\
The eleven-dimensional metric is:
\bq \label{11metric}
\begin{split}
ds^2_{11}& = - \left(\frac{\alpha}{Z_1Z_2Z_3}\right)^{2/3} \left(dt+k\right)^2 + \left(\frac{Z_1Z_2Z_3}{\alpha}\right)^{1/3}ds_4^2 \\
& + \alpha^{2/3}\left(Z_1Z_2Z_3\right)^{1/3}\left[ \frac{dw_1d\bar{w}_1}{Z_1} + \frac{dw_2d\bar{w}_2}{Z_2} + \frac{dw_3d\bar{w}_3}{\alpha Z_3} + \frac{Z_4}{Z_1Z_2}(dw_1d\bar{w}_2 + dw_2d\bar{w}_1) \right]
\end{split}
\eq
where $\alpha = \left(1-\frac{Z_4^2}{Z_1Z_2}\right)^{-1}$ and $ds_4^2$ is a hyper-Kahler base metric.\\

The three-form potential is as follows:
\bq \label{11potential}
\begin{split}
A & = \left(-\frac{\alpha}{Z_1}(dt+k) + B_1\right)\wedge\frac{dw_1\wedge d\bar{w}_1}{-2i} + \left(-\frac{\alpha}{Z_2}(dt+k) + B_2\right)\wedge\frac{dw_2\wedge d\bar{w}_2}{-2i} \\
& \left(-\frac{1}{Z_3}(dt+k) + B_3\right)\wedge\frac{dw_3\wedge d\bar{w}_3}{-2i} + \left(-\frac{\alpha Z_4}{Z_1Z_2}(dt+k) + B_4\right)\wedge\frac{dw_1\wedge d\bar{w}_2 + dw_2\wedge  d\bar{w}_1}{-2i}
\end{split}
\eq
Alternatively we can write:
\bq \label{gaugefields}
A  = A^1\wedge\frac{dw_1\wedge d\bar{w}_1}{-2i} +A^2\wedge\frac{dw_2\wedge d\bar{w}_2}{-2i} + A^3 \wedge\frac{dw_3\wedge d\bar{w}_3}{-2i} + A^4 \wedge\frac{dw_1\wedge d\bar{w}_2 + dw_2\wedge  d\bar{w}_1}{-2i}
\eq
with the obvious definition of the gauge fields $A^I$, $I=1,2,3,4$. Also,
\bq \label{wtori}
w_1=z_2-iz_3 \, , \quad w_2=z_1+iz_4\, , \quad w_3=y+iz
\eq
are the complex torus coordinates.\\

 The BPS equations that need to be solved are:
\bq \label{BPSequations}
\begin{split}
& \Theta_I = *_4\Theta_I \, , \, I=1,2,3,4 \\
& d*_4dZ_1 = - \Theta_2\wedge \Theta_3 \, , \quad d*_4dZ_2=-\Theta_1\wedge \Theta_3 \\
& d*_4dZ_3 = - \Theta_1\wedge \Theta_2 + \Theta_4 \wedge \Theta_4 \, , \quad  d*_4dZ_4 = - \Theta_3 \wedge \Theta_4 \\
& dk + *_4 dk = Z_1\Theta_1 + Z_2\Theta_2 + Z_3 \Theta_3 - 2Z_4\Theta_4
\end{split}
\eq
where $\Theta_I = dB_I$. \\

By choosing a Gibbons-Hawking base metric the general form of the solution \cite{Giusto:2012gt}   is
\bq \label{generalsolution}
\begin{split}
& Z_1 = L_1+\frac{K_2K_3}{V} \, , \quad Z_2 = L_2+\frac{K_1K_3}{V} \\
& Z_3 = L_3+ \frac{K_1K_2-K_4^2}{V} \, , \quad Z_4= L_4+ \frac{K_3 K_4}{V}\\
& k = \left( M + \frac{L_1K_1+L_2K_2 +L_3K_3 -2L_4K_4}{2V} + \frac{(K_1K_2-K_4^2)K_3}{V^2} \right) \left(d\psi +A\right) + \omg \\
& *_3 d\omg = VdM - MdV +\frac{1}{2}\left(\sum_{i=1}^3 K_idL_i - L_i dK_i\right) - \left( K_4 dL_4 - L_4 dK_4\right)
\end{split}
\eq
where we expanded $k=\mu(d\psi+A)+\omg$. Also,
\bq \label{theta}
\Theta^I = \sum_{a=1}^3 \partial_a\left(\frac{K^I}{V}\right)\Omega_+^a \, , \, I=1,2,3,4
\eq
where $\Omega_+^a$ are the Gibbons-Hawking self dual two-forms. Their exact form and more details about the Gibbons-Hawking metrics can be found in the appendix. Closure of the Maxwell fields, $d\Theta^I=0$, implies that the functions, $K^I$ are harmonic:
\bq \label{Kharmonic}
\nabla ^2K^I=0
\eq
%

Thus for the local potential $B_I$ we find:
\bq \label{Bpotential}
B_I = \frac{K_I}{V}(d\psi+A) + \vec{\xi}_I \cdot \vec y
\eq
where $\vec{\nabla}\times \vec{\xi}_I = - \vec{\nabla}K_I$ , $\vec{y}=(x,y,z)$ and $\vec{\nabla}$ is with respect to $\IR^3$.\\

The solution displayed above remains invariant under the following gauge transformations:\\
\bq \label{gaugetrans}
\begin{split}
& K_I \ra K_I + c_I V \,  ,\qquad I=1,2,3,4\\
& L_I \ra L_I - C_{IJK}c^JK^K - \frac{1}{2}C_{IJK}c^Jc^KV + \delta_{I,3}\left(2c_4K_4 + c_4^2V\right) \, , \qquad I,J,K=1,2,3 \\ 
& L_4 \ra L_4 -c_3c_4V - K_3c_4-c_3K_4 \\
& M \ra M - \frac{1}{2}c^IL_I + \frac{1}{12}C_{IJK}\left(c^Ic^Jc^K V + 3c^Ic^JK^K \right) + c_4L_4 - c_3c_4K_4 - \frac{1}{2}c_4^2K_3 - \frac{1}{2}c_4^2c_3V \, , \, I,J,K+1,2,3
\end{split}
\eq
where $C_{IJK}=|\epsilon_{IJK}|$ and $c_I$ are constants. These transformations are a direct generalization of the ones in \cite{Bena:2007kg}.

\subsection{Regularity}

Regularity of the metric (\ref{11metric}) from the tori gives:
\bq \label{torireg}
Z_2Z_3>0 \, , \, Z_1Z_3>0\, , \, Z_1Z_2>Z_4^2 \, , \, Z_4Z_3>0
\eq
These conditions guarantee that $\alpha >0$. \\

By completing the square with respect to $d\psi$ the five-dimensional part of the metric becomes:\\
\bq \label{squaredmetric}
ds_5^2 = \left(\frac{\alpha}{Z_1Z_2Z_3}\right)^{2/3}\frac{I_4}{V^2}\left(d\psi + A - \frac{\mu V^2}{I_4}(dt+\omg)\right)^2 + \left(\frac{\alpha}{Z_1Z_2Z_3}\right)^{-1/3}\frac{V}{I_4}\left(I_4ds_3^2 - (dt+\omg)^2\right)
\eq
where:
\bq \label{I4}
\begin{split}
& I_4=\alpha^{-1}Z_1Z_2Z_3V-\mu^2V^2 \Ra \\
 I_4 &= \frac{1}{2}\sum_{I<J=1}^3 K_IK_JL_IL_J - \frac{1}{4}\sum_{I=1}^3 K_I^2 L_I^2 + V\left(L_1L_2-L_4^2\right)L_3\\
& +\left(K_1L_1+K_2L_2-K_3L_3\right)K_4L_4 - K_4^2L_1L_2 - K_1K_2L_4^2 \\
& - 2M\left(K_1K_2-K_4^2\right)K_3 - MV\left(\sum_{I=1}^3 K_IL_I - 2K_4L_4\right) - M^2V^2
\end{split}
\eq
The function $I_4$ is invariant under the gauge transformations (\ref{gaugetrans}).

The absence of CTC's requires:
\bq \label{noCTCsgeneral}
(Z_1Z_2Z_3)^{1/3}V \ge 0 \, ,  \quad I_4 \ge 0
\eq
Writing $ds_3^2=dr^2+r^2d\theta^2+r^2sin^2\theta d\phi^2$ it can be seen from (\ref{squaredmetric}) that at the poles $\theta=0,\pi$ there is the additional danger of closed timelike curves unless $\omg$ vanishes at these points.

\section{Bubbling}
Our purpose is to resolve the singularity and construct smooth microstate geometries by allowing the multi-centered Gibbons-Hawking metric to be ambipolar. In our analysis we follow the steps of \cite{Bena:2007kg,Bena:2005va}   Specifically for the functions characterizing the solution we write the ansatz:
\bq \label{functions}
\begin{split}
& V = h + \sum_{j=0}^N\frac{q_j}{r_j} \, , \quad 
 L^I = l^I_{\infty} + \sum_{j=0}^N\frac{l_j^I}{r_j}\\
& K^I = k^I_{\infty} + \sum_{j=0}^N\frac{k_j^I}{r_j} \, , \quad
 M = m_{\infty} + \sum_{j=0}^N \frac{m_j}{r_j}
\end{split}
\eq
with $I=1,2,3,4$. We also set as $r_j=|\vec{y}-\vec{y}_j|$ the distance from the $j^{th}$ center located at $\vec{y}_j$. The charges $q_j$ can be positive or negative integers with the requirement that $\sum_{j=0}^N q_j =q_{tot}=1$, for the space to be asymptotically Minkowski.\\

\subsection{Flat Asymptotics}
To have five-dimensional Minkowski spacetime at infinity we need $h=0$, which means that $k_{\infty}^I=0$ as well. We want $\frac{\alpha}{Z_1Z_2Z_3} \ra 1$ and $\mu \ra 0$ as $r\ra \infty$. So,
\bq \label{flatasymptotics}
l_{\infty}^3\left(l_{\infty}^1l_{\infty}^2 - (l_{\infty}^4)^2\right) = 1 \, , \quad m_{\infty} = -\frac{\sum_{j=0}^N\left(\sum_{I=1}^3l_{\infty}^Ik_j^I - 2l_{\infty}^4k_j^4\right)}{2q_{tot}}
\eq
An obvious choice for flat asymptotics is $l_{\infty}^2=l_{\infty}^3=l_{\infty}^4=1$ and $l_{\infty}^1=2$. The choice $l_{\infty}^1=l_{\infty}^2=l_{\infty_4}=3$ and $l_{\infty}^4=0$ might be interesting since it resembles the case in \cite{Bena:2007kg}, but as it can be seen from (\ref{flatasymptotics}) and (\ref{bubble}) it may also disentangle some interesting physical sectors which involve the fourth dipole charge $k_4$.

\subsection{Regularity at the critical surfaces}
Since the solution is ambipolar there are regions where $V<0$ and $V=0$ surfaces. Thus we have to check the metric and the fields of the solution in the neighborhood of $V=0$.\\
The tori warp factors and the function $\alpha$ contain the same power of $Z_I$'s in the numerator and the denominator and thus are regular at $V=0$. It is also easy to check from (\ref{squaredmetric}) that the five-dimensional part of the metric is regular at $V=0$ since $I_4$ , $(Z_1Z_2Z_3)^{1/3}V$ and $\mu V^2$ remain finite. Thus the metric and its inverse are regular at $V=0$.\\ 
For $V<0$ we need again $I_4\ge 0 $. That can be seen from (\ref{squaredmetric}), since if we focus near $V=0$ the warp factor $\left(\frac{\alpha}{Z_1Z_2Z_3}\right)^{-1/3}V$ is $\left((K_1K_2-K_4^2)^2K_3^2\right)^{1/3}$ which is positive.\\
Regarding the other fields of the solution from (\ref{11potential}) and (\ref{generalsolution}) it is simple to see that no pathologies are hidden in $A_I$ and $\omg$ as $V=0$. Actually for the gauge fields we get $A_I \sim 0$ for $I=1,2,3,4$ in the same manner as in \cite{Bena:2007kg}.

\subsection{Regularity at the centers}
We also have to check that the solution is regular as one approaches the Gibbons-Hawking centers $r_j \ra 0$.
In order to cancel the singularities in the functions $Z_I$ we require:
\bq \label{Zregularity}
\begin{split}
& l_j^1 = - \frac{k_j^2k_j^3}{q_j} \, , \quad l_j^2 = - \frac{k_j^1k_j^3}{q_j} \\
& l_j^3 =  \frac{(k_j^4)^2-k_j^1k_j^2}{q_j}  \, , \quad l_j^4 = - \frac{k_j^3k_j^4}{q_j} 
\end{split}
\eq
and cancelling the singularities in $\mu$ requires,
\bq \label{muregularity}
m_j = \frac{k_j^3}{2q_j^2}\left(k_j^1k_j^2-(k_j^4)^2\right)
\eq
The magnetic fields strengths $\Theta_I$ are regular since the singularities of the functions $V$ and $K_I$ coincide. This suggests that if $h=0$ and our base space is flat $\IR^4$ for $\Theta_I$ to be regular we should also have $k_{\infty}^I=0$ , $I=1,2,3,4$. \\
Since $Z_I$ and $\mu$ are finite as $r_j\ra 0$, then from (\ref{11metric}) the absence of closed timelike curves requires $\mu \ra 0$ at this limit. Also from (\ref{BPSequations}) there is the danger of Dirac-strings in $\omg$ as there are $d\frac{1}{r_j}$ terms in the right hand side of the equation. Another way to write the equation for $\omg$ is:
\bq \label{omgequation2}
*_3d\omg=Vd\mu - \mu dV - V\left( \sum_{I=0}^3 Z_I d\left(\frac{K^I}{V}\right) - 2Z_4 d \left(\frac{K^4}{V}\right)\right)
\eq
Also, because $\mu$, $Z_I$ and $K_I/V$ go to finite values at the limit $r_j\ra 0$ the only term that can lead to Dirac strings in (\ref{omgequation2}) is $\mu dV$. Thus the absence of both Dirac strings and closed timelike curves requires that $\mu \ra 0$ as $r_j \ra 0$. From this requirement we take the bubble equations which are necessary integrability conditions for the regularity of the solution.
\bq \label{bubble}
\sum_{j=0 , j\neq i}^N \left( \left(\left(\Pi^1_{ij}\Pi^2_{ij} - (\Pi_{ij}^4)^2\right)\Pi_{ij}^3 \right) \frac{q_iq_j}{r_{ij}}\right)    = - 2m_{\infty}q_i - \sum_{I=1}^3 l_{\infty}^Ik_i^I + 2l_{\infty}^4k_i^4
\eq
where $\Pi_{ij}^I = \frac{k_j^I}{q_j} - \frac{k_i^I}{q_i}$ are the magnetic fluxes running through the two-cycle formed between the centers $i$ and $j$ and $r_{ij}=|\vec{y}_i-\vec{y}_j|$ the distance between them.

\subsection{Solving for $\omg$}

For $\omg$ we find:
\bq \label{omg}
\vec{\omg} = \frac{1}{4} \sum_{i,j=0}^Nq_iq_j\left( \Pi_{ij}^1\Pi_{ij}^2 - (\Pi_{ij}^4)^2\right)\Pi_{ij}^3\vec{\omg}_{ij}
\eq
where:
\bq \label{omgij}
\omg_{ij} = - \frac{x^2+y^2 + (z-a+r_i)(z-b-r_j)}{(a-b)r_ir_j}d\phi_{ij}
\eq
and we have set the z-axis along the two points $i$ and $j$ so that $\vec{y}_i = (0,0,a)$ , $\vec{y}_j = (0,0,b)$ and $a>b$. The angle $\phi_{ij}$ is the azimuthal angle of the $(i,j)$ coordinate system with z-axis passing through the points $i$ and $j$. The functions $\omg_{ij}$ vanish along the z-axis and thus have no dirac string singularities. They satisfy the equation,
\bq \label{omgequation}
\vec{\nabla}\times\vec{\omg}_{ij} = \frac{1}{r_i}\vec{\nabla}\frac{1}{r_j} - \frac{1}{r_j}\vec{\nabla}\frac{1}{r_i} + \frac{1}{r_{ij}}\left(\vec{\nabla}\frac{1}{r_i} - \vec{\nabla}\frac{1}{r_j}\right)
\eq
Taking $\vec{\nabla}\times\vec{\omg}$ and using (\ref{omg}) together with (\ref{omgequation}) and the regularity equations (\ref{bubble}), (\ref{Zregularity}), (\ref{muregularity}) we obtain the right hand side of the last equation in (\ref{generalsolution}).

\subsection{Asymptotic Charges}
The electric charges $\tilde Q_I$ are given by the asymptotic behaviour of the electric potentials $Z_I$ at infinity as follows:
\bq \label{Zexpansion}
Z^I = l_{\infty}^I + \frac{\tilde Q^I}{4r} \, , \quad r\ra \infty
\eq
Thus by expanding $Z_I$ and making use of (\ref{functions}) and (\ref{Zregularity}) we get:
\bq \label{electriccharges}
\begin{split}
& \tilde Q^1= - 4\sum_{j=0}^N \frac{\tilde{k}_j^2\tilde{k}_j^3}{q_j} \, , \quad \tilde Q^2= - 4\sum_{j=0}^N \frac{\tilde{k}_j^1\tilde{k}_j^3}{q_j} \\
& \tilde Q^3= - 4\sum_{j=0}^N  \frac{\tilde{k}_j^1\tilde{k}_j^2 - (\tilde{k}_j^4)^2}{q_j} \, , \quad \tilde Q^4= - 4\sum_{j=0}^N \frac{\tilde{k}_j^3\tilde{k}_j^4}{q_j}
\end{split}
\eq
where the quantities $\tilde{k}_j^I$ are invariant under the gauge transformations (\ref{gaugetrans}),
\bq \label{ktilde}
\tilde{k}_j^I = k_j^I - q_j\sum_{i=0}^N k_i^I \, , \quad I=1,2,3,4
\eq
As it was mentioned in \cite{Giusto:2012gt} the D1-D5 CFT can be obtained by setting $\tilde Q_4=0$. It can be seen from (\ref{electriccharges}) there is a variety of possibilities of achieving that without setting $k_j^4=0$. Each one of these choices though should be checked for consistency with the regularity constraints of the solution. Studies of the bubbled equations of the STU model suggest when one of the asymptotic charges is being set to zero the solution becomes pathological. However, the extra freedom of parameters of this model may allow such a choice.\\ 

The angular momentum can be derived from the expansion
\bq \label{muexpansion}
k \sim \frac{1}{16r}\left( (J_1+J_2) + (J_1-J_2)cos\theta\right) d\psi
\eq
and we get:
\bq \label{JR}
J_R=J_1+J_2 = 8\sum_{j=0}^N \frac{(k_j^1k_j^2 - (k_j^4)^2)k_j^3}{q_j^2}
\eq
\bq \label{JL}
\vec{J}_L=\vec J_1 - \vec J_2 = \sum_{i,j=0, j\neq i}^N J_{L_{ij}}
\eq
where:
\bq \label{JLij}
J_{L_{ij}}= -8  \left( \left(\left(\Pi^1_{ij}\Pi^2_{ij} - (\Pi_{ij}^4)^2\right)\Pi_{ij}^3 \right)q_iq_j \frac{(\vec y_i - \vec y_j)}{r_{ij}}\right) 
\eq
is the angular momentum flux vector associated with the $ij^{th}$ bubble and in the derivation of it we used (\ref{bubble}). Comparing with the already known bubbled geometries \cite{Bena:2007kg,Berglund:2005vb} we see that most our results can be obtained from the old ones by making the substitution $\Pi^1_{ij}\Pi^2_{ij} \ra \Pi^1_{ij}\Pi^2_{ij} - (\Pi_{ij}^4)^2$ and $k_1k_2 \ra k_1k_2-k_4^2$. There is no need so make such a substitution for the electric  charges as they have been dissolved into the magnetic fluxes (\ref{Zregularity}).

\section{A three charge supertube in IIB}

We start with a solution with three electric charges $Q_1$, $Q_2$, $Q_4$ and one dipole magnetic charge $k_3$ which we call a three charge supertube. The solution has a tubular shape since it wraps around the Gibbons-Hawking fiber $\psi$. It can be directly obtained from the one in \cite{Giusto:2012gt}  by setting $Q_3=0$ and $k_1=k_2=k_4=0$. For more generality we are going to assume $V=h+\frac{q}{r}$. The functions describing the solution are as follows:
\bq \label{supertube}
\begin{split}
& L_1 = l_{\infty}^1 + \frac{Q_1}{4r_R} \, , \quad L_2=l_{\infty}^2 + \frac{Q_2}{4r_R} \, , \quad L_3 = l_{\infty}^3 \, , \quad L_4 = l_{\infty}^1 + \frac{Q_4}{4r_R} \\
& K_1=K_2=K_4=0 \, , \quad K_3=\frac{k_3}{r_R} \, , \quad M=m_\infty + \frac{m}{r_R}
\end{split}
\eq
where $r_R=\sqrt{r^2+R^2-2rRcos\theta}$ and the supertube is positioned at distance R from the origin along the positive z-axis.\\
Since the electric potentials $Z_I$ are not sourced at the origin, to keep $I_4\ge 0$ as $r\ra 0$ we need $\mu\ra 0$ at this limit. Consequently:
\bq \label{0reg}
m_{\infty}=-\frac{m}{R}
\eq
By canceling the remaining terms in the right hand side of the $\omg$ equation in (\ref{generalsolution}) so that it looks like (\ref{omgequation}) we get the condition for the absence of closed timelike curves:
\bq \label{CTC}
2mV_R=l_{\infty}^3k_3
\eq
where $V_R=\left(h+\frac{q}{R}\right)$.\\

The metric of the four charge system in the IIB frame is:
\bq \label{IIBmetric}
ds^2= \frac{\alpha}{\sqrt{Z_1Z_2}}\left(-\frac{1}{Z_3}(dt+k)^2 + Z_3\left(dy - \frac{dt+k}{Z_3} + B_3\right)^2\right) + \sqrt{Z_1Z_2}ds_4^2 + \sqrt{\frac{Z_1}{Z_2}}ds^2_{T^4}
\eq
The ten-dimensional metric is split between between a six-dimensional part and four dimensions compactified in $T^4$. The six-dimensional metric can be obtained from the five-dimensional one by promoting one of the gauge fields to a Kaluza-Klein coordinate. This field is $A^3$ in our case and when (\ref{streg}) holds, resolves the singularity in five dimensions.
To examine regularity along the supertube as $r_R \ra 0$ we look for potential singularities along the fiber. Thus collecting all the $(d\psi + A)^2$ terms from (\ref{IIBmetric}) we obtain:
\bq \label{fiberterms}
\frac{\alpha}{V^2\sqrt{Z_1Z_2}}\left(\frac{Z_1Z_2V}{\alpha} - 2\mu VK_3+Z_3K_3^2\right)
\eq
thus for regularity as $r_R \ra 0$  and from \ref{CTC} we need:
\bq \label{streg}
m=\frac{Q_1Q_2-Q_4^2}{32k_3}
\eq
Equation (\ref{streg}) together with (\ref{CTC}) fixes the location of the supertube in terms of its electric and dipole magnetic charges. Once again, by substituting in (\ref{streg}) the combination $Q_1Q_2-Q_4^2\ra Q_1Q_2$ we get the regularity condition the STU model supertube. Since we have already set $k_4=0$ to get a supertube dual to a D1-D5 CFT state we need $Q_4=0$ which takes us back to the STU model supertube. Supertubes correspond to unbound states in the dual CFT. Thus probably only bound states in the new CFT may lead us back to the D1-D5 sector. It would be interesting to find the combinations of generalized supertubes that achieve the latter. 

\section{Concluding Remarks}
Using the M-theory framework we have obtained microstate geometries corresponding to the black rings presented in \cite{Giusto:2012gt}. The off-diagonal term in the supergravity gauge field gives an additional flux, $\Pi^4_{ij}$, in the bubble equations, which dissolves the fourth electric charge and resolves the singularity associated with it.  A smooth supertube solution with three electric charges and one dipole magnetic charge has been shown to exist in this framework as well. Most of the old regularity equations of the STU model can be rewritten for the case of four charges by replacing the quadratic $X_1X_2\ra X_1X_2-X_4^2$ where $X_I$ some parameters of the solution. It is interesting that the fourth flux, $\Pi^4_{ij}$, couples only to $\Pi_{ij}^3$. This reflects the fact that even in the M-theory frame the extra $U(1)$ gauge field is not in equal footing with the previous three. This should be related to the geometry being 1/8-BPS in spite of having four charges. The quadratic $Q_X \equiv X_1X_2-X_4^2$ also appears in \cite{Giusto:2012gt} and originates from the intersection numbers $C_{IJK}$ which occur after truncating eleven-dimensional supergravity down to five dimensions. We have,
\bq \label{intersection}
C_{IJK}=|\epsilon_{IJK}|\, \, , \, \, I,J,K=1,2,3 \quad , \quad C_{344}=-2
\eq
with all the rest being zero. The cubic invariant factorizes into the quadratic $Q_X$ as:
\bq \label{volume}
\frac{1}{6}C_{IJK}X^IX^JX^K =\left( X^1X^2-(X^4)^2\right)X^3 = Q_XX^3
\eq
The latter constraint defines the symmetric space $SO(1,1) \otimes \left(SO(1,2)/SO(2) \right)$. This space is one of the many possible truncation of eleven dimensional supergravity to five dimensions with $\mathcal{N}=2$ supersymmetry \cite{Cadavid:1995bk,Papadopoulos:1995da}. By exploring further down this road, there may be more general families of 1/8-BPS black hole hair which allow further generalizations of the quadratic in terms of cubic and other symplectic invariants.\\

 The algebraic similarities with the STU case are so many that it would be straightforward to perform the analysis done for the BPS case in \cite{Bena:2008nh,Bena:2008wt} and for the non-BPS case and in \cite{Bena:2009en,Vasilakis:2011ki}. The three charge supertube solution we presented makes use of the field $A^3$ as a Kaluza-Klein coordinate to oxidize the five-dimensional metric. Because of the symmetry that exists in the STU model it is trivial to oxidize the metric with any of the gauge fields $A^I$. We can try doing the same in this case and then find a connection between generalized supertubes and bubbled geometries by using spectral flow transformations \cite{Bena:2008wt} or some generalized version of them. This could lead to a larger family of microstate geometries being constructed. We plan exploring the latter in future work. Then one can see how much entropy these solutions take by putting supertubes in an ambipolar base spaces and exploring the entropy enhancement mechanism \cite{Bena:2008nh}. For the non-BPS case we can break supersymmetry by reversing the holonomy of the background with respect to the duality of the magnetic field strengths $\Theta^I$ and construct multicenter non-BPS solutions \cite{Bena:2009en}. Then in the spirit of \cite{Vasilakis:2011ki} it would be interesting to examine how the fourth charge affects the tolerance of the non-BPS microstate solution to supersymmetry breaking. There might be the case that one can use the fourth dipole charge to dilute the holonomy of the background, which breaks the supersymmetry, while keeping the values of the other electric and magnetic charges in a region that was previously excluded. \\
 
 Finally, exploring the work of \cite{Giusto:2011fy} it would be interesting to make the connection between the microstates at strong gravitational coupling and the states of the dual CFT. The four charge solutions are dual to a CFT with central charge \cite{Giusto:2012gt} :
 \bq \label{central}
 c \sim \tilde Q_1 \tilde Q_2 - \tilde Q_4^2
 \eq
 This CFT is still unknown and we believe that this and subsequent work may shed more light towards its nature.

\bigskip
\bigskip
\bigskip
\leftline{\bf Acknowledgements}
\smallskip
 I would like to thank Nicholas Warner for valuable discussions. OV would like to thank the USC Dana and David Dornsife College of Letters, Arts and Sciences for support through the College Doctoral Fellowship and the USC Graduate School for support through the Myronis Fellowship. This work of is supported in part by DOE grant DE-FG03-84ER-40168.


\appendix 

\section{Gibbons - Hawking metrics}
\label{appendixA}
\renewcommand{\theequation}{A.\arabic{equation}}
\renewcommand{\thetable}{A.\arabic{table}}
\setcounter{equation}{0}
Gibbons-Hawking metrics have the form,
\begin{equation}
ds^2_4 ~=~  V^{-1}(d\psi + A)^2 ~+~  V \, (dx^2 + dy^2 +   dz^2)     \,,
\label{GHmet}
\end{equation}
with 
\begin{equation}
 V ~=~  h ~+~ \sum_{i=0}^N {q_i \over r_i}   \,,
\label{TNVdefn}
\end{equation}
where $r^2 \equiv \vec y \cdot \vec y$ with  $\vec y \equiv (x,y,z)$. The distance of the $i^{th}$ Gibbons-Hawking center located at $\vec{y}_i$ from the origin is $r_i=|\vec{y}-\vec{y}_i|$. \\
The metric is hyper-K\"ahler if 
\begin{equation}
 \vec \nabla V ~=~ \pm  \vec \nabla \times \vec A   \,,
\label{VAreln}
\end{equation}
Some common examples are when $V=\frac{1}{r}$  and $V=h+\frac{q}{r}$ where the space is $\IR^4$ and Taub-NUT respectively. Introducing frames
\begin{equation}
\hat e^1~=~ V^{-{1\over 2}}\, (d\psi ~+~ A) \,,
\qquad \hat e^{a+1} ~=~ V^{1\over 2}\, dy^a \,, \quad a=1,2,3 \,,
\label{GHframes}
\end{equation}
we can write a natural base of self-dual and anti-self-dual two-forms,
\begin{equation}
\Omega_\pm^{(a)} ~\equiv~ \hat e^1  \wedge \hat
e^{a+1} ~\pm~ \coeff{1}{2}\, \epsilon_{abc}\,\hat e^{b+1}  \wedge \hat e^{c+1} \,, \qquad a =1,2,3\,.\
\label{twoforms}
\end{equation}
%


\end{document}